%% file: imagmpsdmft.tex
\documentclass[aps,pra,twocolumn,nobalancelastpage,footinbib,
noshowpacs,preprintnumbers,superscriptaddress]{revtex4-1}
\usepackage{amsmath,amssymb,amsfonts,fixmath,siunitx}
\usepackage{bm,graphicx,color}
\usepackage[pdftex,colorlinks=true,bookmarksnumbered,bookmarksopen=true]{hyperref}

\include{latexCommandDef}

\def\figdir{./}
\def\numB{0.4}
\def\numC{0.22}

\begin{document}

\title{Imaginary-time matrix product state impurity solver for dynamical mean-field theory}

\author{F. Alexander Wolf}
\affiliation{
Department of Physics, Arnold Sommerfeld Center for Theoretical Physics,
LMU Munich,
80333 M\"unchen, Germany}
\author{Ara Go}
\affiliation{Department of Physics, Columbia University in the City of New York, New York, NY 10027, USA}
\author{Ian P. McCulloch}
\affiliation{
Centre for Engineered Quantum Systems,
School of Physical Sciences,
The University of Queensland,
Brisbane, Queensland 4072, Australia}
\author{Andrew J. Millis}
\affiliation{Department of Physics, Columbia University in the City of New York, New York, NY 10027, USA}
\author{Ulrich Schollw\"ock}
\affiliation{
Department of Physics, Arnold Sommerfeld Center for Theoretical Physics,
LMU Munich,
80333 M\"unchen, Germany}

\date{\today}

\begin{abstract}
We present a new impurity solver for dynamical mean-field theory based on imaginary-time evolution of matrix product states. This converges the self-consistency loop on the imaginary-frequency axis and obtains real-frequency information in a final real-time evolution. Relative to computations on the real-frequency axis, required bath sizes  are much smaller and less entanglement is generated, so much larger systems can be studied. The power of the method is demonstrated by solutions of a three band model in the single and two-site dynamical mean-field approximation. Technical issues are discussed, including details of the method, efficiency as compared to other matrix product state based impurity solvers, bath construction and its relation to real-frequency computations and the analytic continuation problem of quantum Monte Carlo, the choice of basis in dynamical cluster approximation, and perspectives for off-diagonal hybridization functions.
\end{abstract}

\maketitle

\section{Introduction}

Dynamical mean-field theory (DMFT) in its single-site \cite{metzner89,georges92,georges96} 
and cluster \cite{maier05,kotliar06} variants is among the most widely employed 
computational techniques for solving quantum many-body problems. 
At the core of a numerical solution of DMFT is an \tit{impurity solver}: an algorithm 
for solving a quantum impurity problem. The most prominent examples of impurity solvers 
are continuous time quantum Monte Carlo (CTQMC) \cp{rubtsov05,werner06,gull11}, exact diagonalization (ED) \cp{caffarel94,capone07,liebsch12}, the numerical renormalization group (NRG) \cp{bulla08},  
and the density matrix renormalization group (DMRG) \cp{garcia04}. While all methods 
have their strengths, key limitations mean that fundamentally important classes of problems 
have not yet been adequately addressed. Other recent suggestions for impurity solvers \cp{li15,wang15,arsenault15,schuler15,granath12,shinaoka14} including in particular  the 
computationally inexpensive density matrix embedding theory (DMET) \cp{knizia12}, 
are promising but have not been tested in detail.

CTQMC is widely employed but its application to situations involving low point symmetry,  non-Hubbard interactions or multiple relevant orbitals is limited by the fermionic sign problem. Reaching low temperatures becomes highly computationally expensive while calculating real-frequency  information requires analytical continuation, a numerically ill-posed procedure fraught with practical difficulties. 

ED makes no assumption on the interaction and  does not have a sign problem. It is limited  by the size of the Hilbert space that can be studied, meaning in practice that it is restricted  to a small number of correlated sites to which only a small number of bath sites can be attached. Recently, improvements have been achieved by  considering only restricted subspaces of the Hilbert space \cp{lu14,zgid12,lin13i,lin13ii}, but the size of problem remains a significant limitation.

NRG converges the DMFT loop on the real-frequency axis  and very effectively obtains real-frequency information in the low frequency limit.
Current applications have been to relatively small problems (the most recent achievement is a solution of the single-site DMFT approximation to  a three-band model \cp{stadler15}) and it remains to be seen how far the method can be extended.

DMRG \cp{white92} is a set of algorithms operating on the space of matrix product states (MPS) \cp{schollwock11}.  It has been found to be extremely powerful for the calculation of ground states of  one-dimensional quantum systems \cp{schollwock05,schollwock11}; it was very successfully extended to the calculation of spectral functions which, in contrast to NRG,  it obtains with equal resolution across the spectrum, see \eg \cp{holzner11,wolf15}. 
In pioneering work the method was applied as a DMFT solver by \ct*{garcia04} and \ct*{nishimoto04i} with important further work done by these and other authors \cp{karski05,nishimoto06,garcia07,karski08,peters11,ganahl14,ganahl14i}. However, the method has not been widely accepted, perhaps because high-quality data were presented only for the single-site approximation to the single-band Hubbard model. Recently the method was used to reliably solve a two-site DCA \cp{wolf14}, and  insights into the entanglement of the impurity problem made it even more powerful \cp{wolf14i}. In view of these advances, DMRG now is a candidate  for a highly flexible low-cost impurity solver, which can in addition be efficiently employed  in the non-equilibrium formulation of DMFT \cp{gramsch13,wolf14i,balzer15}.

This paper reformulates the DMRG method as an  impurity solver for DMFT on the imaginary-time axis (\secref{secMethod}). As we will show, this strongly reduces entanglement  and necessary bath sizes. The price to be paid is a reduced resolution on the real-frequency axis, which we study in detail by comparing with calculations that converge the DMFT loop on the real-frequency axis (\secref{secBath}). The reformulation allows a much larger class of problems to be addressed, including some that are unreachable by other methods, due e.g.~to the sign problem, the size of the correlated cluster or the number of bands. We illustrate this with calculations for three band models in the single-site and two-site DMFT approximation (\secref{secChecks}). We append discussions of the optimization of typical DMFT Hamiltonians (\appref{secTech}) and of the entanglement in different  representations of the DCA including a discussion of off-diagonal hybridization functions (\appref{secBasis}).

\section{Method}
\label{secMethod}
\subsection{Overview: Green's functions in DMRG}

The computational key challenge in DMFT is the computation of 
the full frequency dependence of the Green's function of a quantum impurity  
model involving an essentially arbitrary bath. The ``size'' (number of correlated sites $L_c$) 
of the impurity model should be as large as possible and the kinds of interaction that can be 
treated should be as general as possible. The Green's function is used in a self-consistency loop, 
which may require many iterations for convergence. The solution should be as inexpensive 
as feasible, and must run automatically, without need for manual optimization of parameters 
or procedures. In this subsection we present a qualitative discussion of the issues involved 
in computing the Green's function using DMRG methods, to motivate the work described in detail below. 

Within DMRG one computes Green's functions by first representing 
the system ground state $\ket{E_0}$ as an MPS. One then generates a one electron 
(one hole) excitation $\ket{\psi^>_0}=d\dag\ket{E_0}$ $\left(\ket{\psi^<_0}=d\ket{E_0}\right)$ 
by applying a creation (annihilation) operator $d\dag$ ($d$) to $\ket{E_0}$. 
While the state $\ket{\psi_0^\gtrless}$ is at most as entangled as the ground state $\ket{E_0}$ \cp{wolf15}, 
in order to compute a Green's function, one has to perform further operations 
on $\ket{\psi_0^\gtrless}$. These operations typically increase entanglement
and by that the \tit{bond dimension} of an MPS, which ultimately limits all computations.

Let us be more concrete and consider a general MPS of bond dimension $m$ for a system with $L$ sites 
and open boundary conditions. 
Defining $A^{\sigma_i}, B^{\sigma_i} \in \mathbb{C}^{m\times m}$ 
for $i\neq 1,L$ and $A^{\sigma_1} \in \mathbb{C}^{1\times m}$, $B^{\sigma_L} \in \mathbb{C}^{m\times 1}$, 
where $\sigma_i \in \{0,\uparrow,\downarrow,\uparrow\downarrow\}$ labels a local basis state of the Hilbert space, 
any MPS can be represented as \cp{schollwock11}
\eq{\label{MPS}
\ket{\psi\tl{MPS}} =\!\! \sum_{\sigma_1,...,\sigma_L} \!\! A^{\sigma_1}...\,A^{\sigma_{l}} S B^{\sigma_{l+1}} ...\,B^{\sigma_L} \ket{\sigma_1,...,\sigma_L}, 
}
where $S=\tx{diag}(s_1,...,s_m)$ is a diagonal matrix, and $A^{\sigma_i}$ are \tit{left-normalized} and $B^{\sigma_i}$ are \tit{right-normalized}, respectively:
\eq{
\sum_{\sigma_i} A^{\sigma_i \dagger} A^{\sigma_i}  = I ,  \qquad \sum_{\sigma_i} B^{\sigma_i} B^{\sigma_i\dagger}   = I. 
}
Here, $I$ are identity matrices. Left- and right-normalization make \Eqref{MPS} the Schmidt decomposition of $\ket{\psi\tl{MPS}}$
that is associated with partitioning the system at bond $(l,l+1)$. The bond entanglement entropy for the associated 
reduced density matrix can therefore simply be read of from \Eqref{MPS} \cp{schollwock11}
\eq{
S_{(l,l+1)}^{\tx{ent}} = \sum_{\nu=1}^m s_\nu^2 \ln s_\nu^2.
}
When subsequently we refer to an \tit{entanglement growth} associated with repeated 
operations on $\ket{\psi\tl{MPS}}$ this implies the need to adjust the bond dimension $m$  
such that $\ket{\psi\tl{MPS}}$ still faithfully represents a physical state.
If entanglement in the physical state becomes too large, we have to choose $m$ so large that computations with 
MPS become impractical.

Since the first suggestion for computing spectral functions within DMRG \cp{hallberg95}, 
the field has evolved by the important development of the correction vector method \cp{kuhner99,jeckelmann02}. 
The subsequent understanding  of the connection between DMRG and MPS \cp{schollwock11} 
opened the door to many further approaches to computing spectral and Green's functions, 
in particular, time evolution and subsequent Fourier transform \cp{white04,white08},  
an improved Lanczos algorithm \cp{dargel12}, and the Chebyshev recursion \cp{holzner11,braun14,wolf15}.
All of these are formulated for the calculation of spectral functions at $T=0$,
as considered in the present paper, 
and came at much cheaper computational cost 
than the correction vector method \cp{holzner11,wolf15}.
We note that for $T>0$, there are perspectives for even more powerful algorithms:  
it was recently demonstrated that the numerically exact
spectral function of a molecule consisting of several hundreds of interacting spins
could be computed \cp{savostyanov14}. 

These developments (see \appref{secSpec} for more details) 
make MPS-based solvers an attractive possibility for dynamical mean-field theory. 
However, the growth of entanglement arising in all calculations of Green's function  
has limited the systems sizes that have been addressed to date. 
Also, in MPS computations manual 
adjustments, for example choosing optimal broadening \cite{holzner11} or combining 
results of different systems sizes \cp{dargel12}, are still common practice. 
In the rest of this section, we show that these problems can to a large degree be circumvented 
by computing Matsubara Green's functions using imaginary-time evolution.
The imaginary-time framework naturally extends existing approaches   
based on real-time evolution \cp{ganahl14i,wolf14i},
which have recently been shown to provide currently the most efficient 
algorithmic approach to compute real-frequency spectral functions \cp{wolf15}. 

\subsection{Imaginary-time computation}

The central objects of technical interest in this paper are the \tit{greater} and the \tit{lesser} correlation functions $\wt {G}^\gtrless$, which we define for imaginary time $\tau$ 
\begin{subequations}
\label{Ggtr}
\eq{  
\wt G^\gtrless(\tau)  & = \bra{\psi_0^\gtrless} e^{\mp(H-E_0)\tau} \ket{\psi_0^\gtrless} \label{Ggtrlssimag},\\
\wt G^\gtrless(it)  & = \bra{\psi_0^\gtrless} e^{\mp i(H-E_0)t} \ket{\psi_0^\gtrless} \label{Ggtrlssreal}.
}
\end{subequations}
In the second line, we evaluate $\wt G^\gtrless(\tau)\vert_{\tau=it}$ and by that obtain a correlation function for real time $t$,
which will be useful later on. The functions $\wt G^\gtrless$ carry spin and orbital indices associated with 
the spin and orbital indices of the single-particle (hole) excitation $\ket{\psi_0^\gtrless}$, 
but these indices are not explicitly written here. We will discuss the relationship of 
$\wt G^\gtrless$ to the physical Green's functions (which we denote by $G$) below. 

\begin{figure}
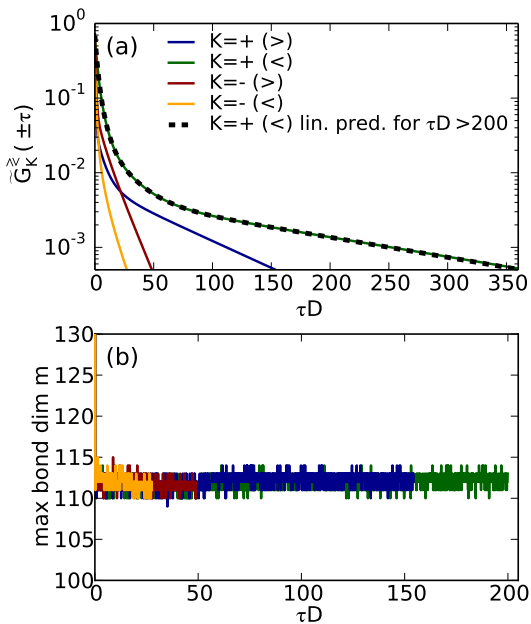

\ig{\numB}{pdf}{\figdir 05b_ferrero_ImagTimeEvol_LinPred}
\ig{\numB}{pdf}{\figdir 05b_ferrero_ImagTimeEvol_BondDim}
\caption{(Color online) 
Panel (a): Imaginary time correlation functions $\wt G^\gtrless(\tau)$  defined in \Eqref{Ggtrlssimag} for an impurity model arising in the context of the  two-site dynamical cluster approximation to the  single-band Hubbard model on the square lattice with next-nearest neighbor hopping  $t'/t=0.3$, half-band width $D=4t$, interaction $U=2.5D$ and band filling $n=0.96$ in the paramagnetic phase. See  Ref.~\onlinecite{ferrero09} for definition of the model and the meaning of the orbital (patch) quantum number  ``$K=\pm$". The dashed line is obtained using \tit{linear prediction} for times $\tau D \geq 200$. Panel (b): Maximal bond dimension $m$ of time evolved states. 
The MPS computation used a Hamiltonian representation of the impurity model with $L_c=2$ correlated sites and $L_b=14$ bath sites.  The hybridization function of the impurity model $\Lambda\th{discr}$  was fitted using \Eqref{costFunc} for $\beta\tl{eff}=315/D$ and $\alpha=0$. For the ground state optimization, we enforced a maximal bond dimension of $m=300$. The Krylov time evolution used a time step of $\Delta t=0.1/D$ and allowed for a maximal global truncation error of $10^{-4}$ at each time step adjusting bond dimensions automatically. This leads to an immediate decay of $m$ at $\tau\simeq0$ from $m=300$ down to $m\simeq110$, as seen in panel (b). We used the global SU(2)  symmetry of the Hamiltonian to reach these low values of the bond dimension. }
\label{figImagTimeEvol}
\end{figure}

While it is not essential in principle, we evaluate \Eqref{Ggtr} using a 
Krylov algorithm \cp{hochbruck97}, which represents the time evolution 
operator in a local Krylov space and is able to treat Hamiltonians 
with long-ranged interactions.  Before performing a time evolution 
computation, one has to compute the initial state $\ket{\psi_0^\gtrless}$ 
using an MPS optimization of the ground state.  As impurity models come with open
boundary conditions, this is well suited for DMRG. 
We discuss this optimization for typical DMFT Hamiltonians in \appref{secGS}.

Figure \ref{figImagTimeEvol} presents representative results based on parameters 
obtained from a two-site DMFT solution of the Hubbard model. Figure \ref{figImagTimeEvol}(a) shows the 
time evolution of $\wt G^\gtrless(\tau)$ out to times as long as 350 times the 
basic timescale (inverse half-bandwidth $D$) of the model, which suffices 
to converge $\wt G^\gtrless(\tau)$ to a precision of $5\cdot10^{-4}$.
Figure \ref{figImagTimeEvol}(b) demonstrates the key advantage that makes this computation possible:  
the lack of growth of maximal bond dimensions $m$ with time 
of the associated imaginary-time evolved states $\ket{\psi^\gtrless(\tau)} = e^{-(H-E_0)\tau} \ket{\psi_0^\gtrless}$.
The imaginary time-evolution operator does not create entanglement 
as it projects on the lowly entangled ground state.

Figure \ref{figImagTimeEvol}(a) reveals additional information about the nature and rate of convergence of $\wt G^\gtrless(\tau)$. In the insulating phase, $H$ has a gap and  $\wt G^\gtrless(\tau)$ decays exponentially irrespective of whether one considers  a finite system or the thermodynamic limit. In the metallic phase, $\wt G^\gtrless(\tau)$ decays algebraically in the thermodynamic limit. For a finite system though, there always remains a small gap, and even though the decay resembles an algebraic decay for short times, it always becomes exponential at long times. The exponential decay can be exploited to speed up computations considerably by  a simple technique known as \tit{linear prediction} \cp{numrec07,white08,barthel09,wolf15}. This technique is an efficient formulation of the fitting problem  for the ansatz function $f(\tau)=\sum_n \alpha_n e^{\beta_n \tau}$, $\alpha_n,\beta_n \in \mathbb{C}, \tau\in \mathbb{R}$, which  can then be used to reliably extrapolate functions with an exponentially decaying envelope. This is illustrated by the dashed black line in \figref{figImagTimeEvol}(a), which has been fitted to match $\wt G^\gtrless(\tau)$ for $\tau D\in[150,200]$ and was then extrapolated for higher times. The solid green line, by contrast, is the result of the MPS computation. Agreement can be seen to be perfect. 
\subsection{Physical Green's functions} 

Of particular interest in the rest of this paper are the imaginary time Green's functions $G\th{mat}(\tau)$ defined via
\begin{equation}
G\th{mat}(\tau)  = -\theta(\tau) \wt G^>(\tau) + \theta(-\tau) \wt G^<(\tau),
\label{Gtau}
\end{equation}
whose Fourier transform gives the Matsubara Green's function
\begin{equation}
G\th{mat}(i\omega_n)=\int_{-\infty}^{\infty} d\tau\, e^{i\omega_n \tau}G\th{mat}(\tau),
\label{Gmat}
\end{equation}
at zero temperature, where $\omega_n=(2n+1)\pi/\beta$ and $\beta\ra\infty$. 
We shall also be interested in the retarded real-time Green's function
\begin{equation}
G\th{ret}(t) = -i\theta(t) (\wt G^>(it) + \wt G^<(it)),
\label{Gret}
\end{equation}
from which the retarded frequency-dependent Green's function is obtained as
\begin{equation}
G\th{ret}(\w)  = \int_{-\infty}^{\infty} dt\, e^{i(\omega +i0^+)t}G\th{ret}(t).
\label{Gret}
\end{equation}

In numerical practice, we evaluate the Fourier transforms leading to \eqref{Gmat} and \eqref{Gret} approximately as
\eq{
G\th{mat}(i\w_n) & = -\int_0^{\tau\tl{max}} \!\!\! d\tau\, \wt G^>(\tau) e^{i\w_n t} + \int_{-\tau\tl{max}}^0 \!\!\!\!\!\!  d\tau\, \wt G^<(\tau) e^{i\w_n t}, \non\\
G\th{ret}(\w) & = -i \int_0^{t\tl{max}} \!\!\!  dt\, \big[\wt G^>(it) + \wt G^<(it) \big] e^{i\w t}, \label{GretApprox}
}
with cutoff times $\tau\tl{max}$ and $t\tl{max}$. This approximation is only controlled if we are able to reach long enough times $\tau\tl{max}$ and $t\tl{max}$, such that $\wt G^\gtrless(\tau)$ and $\wt G^\gtrless(it)$ have converged to zero to any desired accuracy. 

In contrast to a computation on the imaginary axis,
reaching arbitrarily long times $t\tl{max}$ on the real axis is prohibited 
by a logarithmic growth of entanglement, which comes with a power-law growth of bond dimensions. 
In addition, finite-size effects are a severe source of errors because the long-time behavior 
is determined by the bath size. For a numerically exact computation, 
one has to choose the system large enough to observe exponential 
``pseudo-convergence" of $\wt G^\gtrless(it)$ to zero before finite-size effects 
are resolved \cp{wolf15}. In the context of the present paper, we deal with small system 
sizes and will never observe pseudo-convergence. 
In particular there is no exponential pseudo-convergence,
so that \tit{linear prediction} cannot be employed \cp{wolf15}. 
Therefore, when computing the real-frequency spectral function
after converging the DMFT loop, one has to use the further approximation of 
\tit{damping} the finite-size effects that emerge at long times by computing, 
instead of $G\th{ret}(\w)$ in \eqref{GretApprox},
\eq{ \label{GretBroad}
G\th{ret}_\eta(\w) & = -i \int_0^{\infty} \!\!\!  dt\, \big[\wt G^>(it) + \wt G^<(it) \big] e^{i\w t} e^{-\frac{\eta^2t^2}{2}},
}
which yields the \tit{broadened} spectral function $A_\eta(\w) = -\frac{1}{\pi} \tx{Im} G\th{ret}_\eta(\w) = \frac{1}{\sqrt{2\pi}\eta} \int d\w' A(\w') e^{-\frac{(\w-\w')^2}{2\eta^2}}$. Instead of a Gaussian damping and broadening, one could also use an exponential damping leading to Lorentzian broadening, which damps out the original time-evolution information more strongly, though. 

Before presenting detailed benchmark results for the solution of DMFT using imaginary-time evolution of MPS,  
let us clarify which price we have to pay for profiting from the great advantage of not facing entanglement growth.
We do this by comparing the imaginary-time approach (itMPS) to approaches that solve the DMFT loop on the
real axis.

\section{Comparison of imaginary-axis with real-axis computations}
\label{secBath}

The \tit{self-consistency} equation in DMFT relates an impurity model specified 
by a hybridization function and a self energy to a lattice model specified by a 
lattice Hamiltonian and the same self energy. We discuss the issues using the example of 
the dynamical cluster approximation to the single-band Hubbard model
\eq{
G_K\th{latt}(z) 
& = \frac{N_c}{N} \sum_{k\in \mathcal{P}_K} \frac{1}{z+\mu-\varepsilon_k-\Sigma_K(z)},  \label{selfCons}\\
& \stackrel{!}{=} \big[z + \mu -\varepsilon_K - \Sigma_K(z) - \Lambda_K(z)\big]^{-1} = G_K\th{imp}(z). \non
}
Here $\varepsilon_k$ denotes the single-particle dispersion of the lattice and $\mu$ is the chemical potential. In the dynamical cluster approximation, the Brillouin zone, consisting in $N$ momentum vectors $k$, is covered by $N_c$ (for single-band $L_c=N_c$) equal-area tiles (patches), labelled here by $P_K$ and the self energy is taken to be piecewise constant, with being a potentially different function of frequency in each tile. The impurity model is specified by the on-site energy $\varepsilon_K$ and the hybridization function $\Lambda_K(z)$, which is to be determined using the fixed point iteration referred to as the \tit{DMFT loop}. The loop works as follows:  make an initial guess for $\Lambda_K(z)$; compute 
$\Sigma_K(z)=\omega+\mu-\varepsilon_K-\Lambda_K(z) - [G_K\th{imp}(z)]^{-1}$, then update $\Lambda_K$ via $\Lambda_K(z)=z+\mu-\varepsilon_K-\Sigma_K(z)-[G_K\th{latt}(z)]^{-1}$ and repeat this procedure until convergence. 

We discuss two aspects of the comparison of real- and imaginary-frequency solutions of the DMFT self-consistency equation \eqref{selfCons}. The first has to do with the number of bath sites needed to obtain a solution of the self-consistency equation. The second is the accuracy to which the spectral functions of physical interest can be reproduced. 

\begin{figure}
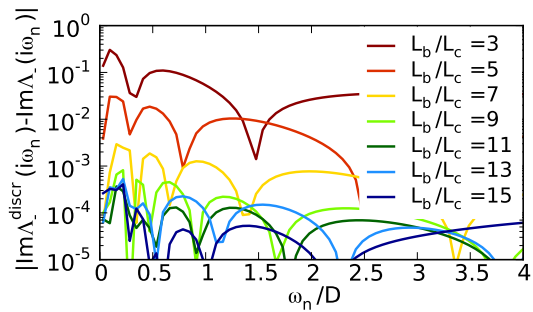

\ig{\numB}{pdf}{\figdir 05d_bathFitNrSitesBeta100}
\caption{(Color online) 
Fit of the hybridization function in the two-site DCA problem
studied in \figsref{figImagTimeEvol} and \ref{figDCA}, but here for
the case $U=0$. The minimization \eqref{costFunc}
was done using $\alpha=0$ and a frequency grid defined by $\beta\tl{eff}=100/D$ and
a cutoff frequency of $\omega_c = 6D$.
Evidently, the quality of the fit does not improve any more for $L_b/L_c\gtrsim9$.
}
\label{figBathFitSites}
\end{figure}

The DMFT self-consistency equation \eqref{selfCons}, defines the hybridization function $\Lambda_K$ as a continuous function in terms of the difference between the computed self energy and the inverse of the lattice Green's function. In DMRG-type methods, the hybridization function $\Lambda_K$ is approximated as the hybridization function $\Lambda\th{discr}_K$ of a discrete impurity model, which is the sum of a number of poles. If the number of poles is too small one cannot construct a meaningful approximation on the real axis \cp{vega15} and a DMFT loop cannot be converged. For this reason DMRG-based solutions of DMFT up to now \cp{garcia04,nishimoto04i,karski05,nishimoto06,garcia07,karski08,peters11,ganahl14,ganahl14i,wolf14,wolf14i}, all of which were real axis computations, have been performed using numbers of bath sites of at least $L_b/L_c \gtrsim 30$, and in the case of the single-band Hubbard model, even much more $L_b/L_c \gtrsim 120$. Use of such a large number of bath sites means that with modest broadening the hybridization function can be reasonably approximated as a continuum, enabling a  stable solution of \Eqref{selfCons}.

By contrast, formulating the problem on the imaginary axis (as is typically done in standard ED solvers where the number of bath sites is strictly limited) automatically smoothens the hybridization function $\Lambda\th{discr}_K$ and permits a stable solution. From the imaginary axis solution one must then determine the discrete set of bath parameters to represent $\Lambda\th{discr}_K$. This is typically done \cp{caffarel94,liebsch12,go15} by numerical minimization of a cost function defined as
\eq{\label{costFunc}
\chi^2 = \frac{1}{N\tl{fit}} \sum_{n=1}^{N\tl{fit}} \w_n^{-\alpha} | \Lambda_K(i\w_n) - \Lambda_K\th{discr}(i\w_n) |^2.
}
Here, $\alpha$ defines a weighting function $\w_n^{-\alpha}$. Choosing $\alpha>0$, \eg $\alpha=1$, attributes more weight to smaller frequencies \cp{senechal10,liebsch12,go15}, which we find helpful when using small bath sizes $L_b/L_c<5$. 
To define the frequency grid for the fit $\w_n=(2n+1)\pi/\beta\tl{eff}$, one defines a fictitious inverse temperature $\beta\tl{eff}$, 
which has no physical significance. We further employ a cutoff frequency $\omega_c$, which implies a finite number $N\tl{fit}$ of fitted Matsubara frequencies.

If one tries to define an analogous cost function for the real axis, the result is useless as then $\Lambda_K\th{discr}(\w+i0^+)$ is a sum of poles,   whereas the hybridization function $\Lambda_K(\w+i0^+)$, as encountered in \eqref{selfCons}, is continuous \cp{vega15}. One can overcome this problem only when using a Lindbladt formalism \cp{dorda14}, which increases the complexity of the problem substantially. 

\begin{figure}
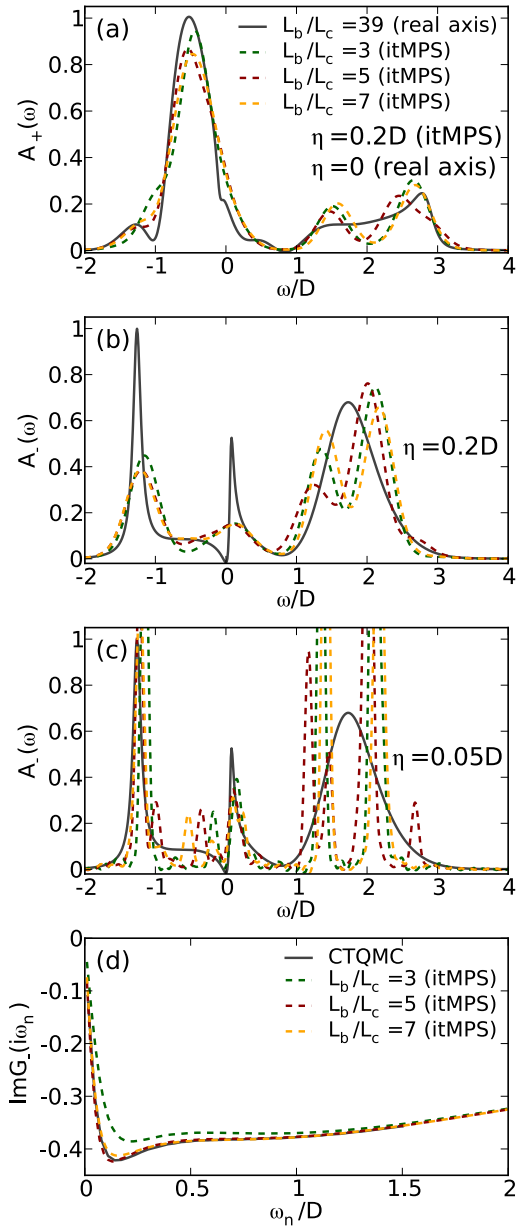

\ig{\numB}{pdf}{\figdir 05c_ferrero_Grealn0.96new}
\ig{\numB}{pdf}{\figdir 05c_ferrero_GrealP2n0.96new}
\ig{\numB}{pdf}{\figdir 05c_ferrero_GrealP2n0.960.05new}
\ig{\numB}{pdf}{\figdir 05c_ferrero_GmatImagP2n0.96}
\caption{(Color online) 
Real- and imaginary-frequency Green's functions computed by converging the DMFT self-consistency equation \Eqref{selfCons} for the two-site dynamical cluster approximation to the single-band Hubbard model on the square lattice with next-nearest neighbor hopping  $t'/t=0.3$, half-band width $D=4t$, interaction  $U=2.5D$, and band filling $n=0.96$ in the paramagnetic phase (as in \figref{figImagTimeEvol}). See Ref.~\onlinecite{ferrero09} for definition of the model and the meaning of the orbital (patch) quantum number  ``$K=\pm$". Panels (a-c): Electron spectral function $A(\omega)=-\frac{1}{\pi}\text{Im}G\th{ret}(\omega)$ obtained by converging on imaginary-frequency axis (itMPS) using number of bath sites and broadenings as specified in the figure, and compared to unbroadened ($\eta=0$) real-frequency axis computation using $L_b/L_c = 39$ bath sites per correlated site of  \oc{wolf14}. Panel (d): Converged  Matsubara Green's function for number of bath sites shown, compared to  numerically exact quantum Monte-Carlo  result of \oc{ferrero09}, computed at $\beta=200/D$.
}
\label{figDCA}
\end{figure}

The minimization of \eqref{costFunc} is done using 
using standard numerical optimization.
The optimization in the initial DMFT
iteration should be done using a global optimization scheme \cp{wales97},
and in subsequent iterations using a local optimization scheme (\eg conjugate gradient), which takes  as
an initial guess for the new bath parameters the values of the previous iteration.
Figure \ref{figBathFitSites} shows the convergence of the fit of the hybridization 
function with the number of bath sites $L_b/L_c$. 
For $L_b/L_c = 7$ one already obtains errors as little as $\simeq 10^{-3}$
and for values $L_b/L_c \gtrsim 9$ the quality of the fit already stops improving.
It is at this point, where we (and all ED-like techniques) 
face the problem of ``analytic continuation" encountered in CTQMC,
namely that Green's functions on the imaginary axis encode
information in a much less usable form than on the real axis.

Consider again the example of the two-site DCA for the single-band 
Hubbard model on the square lattice. In \oc{wolf14}, this problem has been solved 
entirely on the real axis using $L_b/L_c=39$ bath orbitals.
Here, we converge the DMFT loop on the imaginary axis and compute
the spectral function in a final real-time evolution using $L_b/L_c=3,5,7$ bath orbitals.
We compare both solutions in \figref{figDCA}. 
Whereas for the (central) momentum patch ``$+$" shown in \figref{figDCA}(a), we find
satisfactory agreement of the imaginary-axis with the real-axis calculation, 
this is not the case for the (outer) momentum patch ``$-$" shown in \figref{figDCA}(b), 
even though the corresponding imaginary-axis Green's function is well reproduced, see \figref{figDCA}(d). 
Evidently, in \figref{figDCA}(b), the central peak and the pseudo-gap
at the Fermi edge are smeared out by a broadening $\eta=0.2D$ that hides 
finite-size effects to a large degree. Reducing the broadening to 
$\eta=0.05D$ as shown in \figref{figDCA}(c), 
again reveals the pseudo-gap and the central peak; but together with 
unphysical finite-size effects. We observe that the nature of these finite-size effects 
is qualitatively comparable when using different numbers of bath sites $L_b/L_c=3,5,7$. 
Already for $L_b/L_c=3$, we obtain reasonable results. On the imaginary axis, by contrast,  
$L_b/L_c=5,7$ still improve over $L_b/L_c=3$ and almost agree with the numerically
exact QMC data for $\beta=200/D$ of \oc{ferrero09}, see \figref{figDCA}(d). 
However we emphasize that even with the modest number of bath sites used 
here the basic features of the spectral function are reproduced (for example the areas in given frequency ranges).

\section{Three-band calculations}
\label{secChecks}

\subsection{Three-band model in single-site DMFT}
\label{sec3bd1s}

We now demonstrate the power of the method by applying it to 
three-band problems in the single-site approximation 
(where comparison to existing calculations can be made) 
and the two-site approximation. Both was hitherto not accessible 
to DMFT+DMRG computations.

We study the three-band Hubbard-Kanamori model with Hamiltonian (omitting the site index $i$ in the following definition of $H_{\tx{loc},i}$)
\eq{ 
H &=\sum_{k, a,b,\sigma} \varepsilon^{ab}_k d^\dagger_{k,a,\sigma}d_{k,b,\sigma} + \sum_i H_{\text{loc},i}\non\\
H_\text{loc}
& = -\sum_{a, \sigma}(\mu-\Delta_a) n_{a,\sigma} \non \\
& \quad  + \sum_{a} U n_{a,\uparrow} n_{a,\downarrow}                                    \label{HhubkanLat}  \\
& \quad +\sum_{a>b,\sigma} \Big[U' n_{a,\sigma} n_{b,-\sigma} +  (U'-J) n_{a,\sigma}n_{b,\sigma}\Big]  \non\\
& \quad -\sum_{a\ne b}J(d^\dagger_{a,\downarrow}d^\dagger_{b,\uparrow}d_{b,\downarrow}d_{a,\uparrow}
+ d^\dagger_{b,\uparrow}d^\dagger_{b,\downarrow}d_{a,\uparrow}d_{a,\downarrow} + \tx{h.c.}), \non
}
where $i$ labels sites in a lattice and $k$ wave vectors in the first Brillouin zone, $n_{i,a,\sigma}=d^\dagger_{i,a,\sigma}d_{i,a,\sigma}$ is the density of electrons of spin $\sigma$ in orbital $a$  on site $i$, $\mu$ is the chemical potential, $\Delta_a$ is a level shift for orbital $a$, $\varepsilon^{ab}_k$ is the band dispersion, $U$ is the intra-orbital and $U'$ the inter-orbital Coulomb interaction, and $J$ is the  coefficient of the Hund coupling and pair-hopping terms. We adopt the conventional choice of parameters,   $U'=U-2J$, which follows from symmetry considerations  for $d$-orbitals in free space and holds (at least for reasonably symmetric situations) for the $t_{2g}$ manifold in solids \cp{georges12}.
\begin{figure}
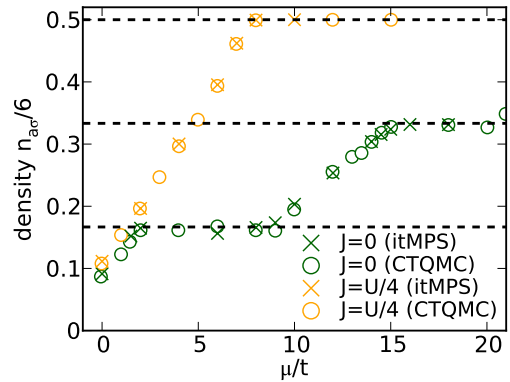

\ig{\numB}{pdf}{\figdir 00a_3band_mu}
\caption{(Color online) 
Density per orbital as function of chemical potential for  three-band Hubbard-Kanamori model \Eqref{HhubkanLat} using the semi-elliptic density of states \Eqref{DOS} and $U=12t$, obtained from single-site DMFT approximation evaluated using imaginary MPS (crosses) and CTQMC data (circles, Figure 1 of \oc{werner09}, inverse temperature $\beta =50/t$). In the DMRG computations the bath fitting was performed using $\beta\tl{eff}=100/t$, $\w_c=6t$ and $\alpha=1$ with three bath sites per correlated site ($L_b/L_c=3$). The maximal matrix dimensions was $m=300$ for the ground state calculation, exploiting the SU(2) symmetry, which leads to the high precision $\expec{(H-E)^2} \simeq 10^{-14}$. For the time evolution, we computed $\wt G_{a}^{\gtrless}(\tau)$ in \eqref{Ggtrlssimag} in steps of $\Delta \tau=0.1/t$ allowing a global truncation error of $5\cdot10^{-4}$ per step, up to imaginary time $\tau\tl{max}=100/t$ and used \tit{linear prediction} for higher times. 
}
\label{fig3bdMIT}
\end{figure}

We study the orbital-diagonal and orbital-degenerate case ($\Delta_a=0$) on the Bethe lattice, \ie the non-interacting density of states is semi-elliptic, 
\eq{\label{DOS}
A_{a,0}(\omega) = \frac{1}{\pi t} \sqrt{1 - (\tfrac{\omega}{2 t})^2}. 
}
In the single-site approximation, the impurity Hamiltonian used within DMFT is given by
\eq{  \label{Hhubkan}
H 
&= H_\text{loc}+H_\text{coupl}+H_\text{bath}, \non\\
H_\text{coupl}
&=\sum_{l,a,\sigma} V_{l,a,\sigma} ~ d^\dagger_{a,\sigma}c_{l,a,\sigma}+\tx{h.c.}, \\
H_\text{bath}
&=\sum_{l,a,\sigma} \varepsilon_{l,a.\sigma} ~ c^\dagger_{l,a,\sigma}c_{l,a,\sigma}, \non
}
where $c^\dagger_{l,a,\sigma}$ creates a fermion in the bath orbital $l$, $V_{l,a,\sigma}$ describes the coupling
of the impurity to the orbital $l$, and $\varepsilon_{l,a.\sigma}$ denotes the potential energy of orbital $l$.
The hybridization function is then given by
\eq{
\Lambda_{a,\sigma}\th{discr}(z) = \sum_{l=1}^{L_b/L_c} \frac{|V_{l,a,\sigma}|^2}{z-\varepsilon_{l,a.\sigma}}.
}

Figure \ref{fig3bdMIT} compares the dependence of the particle density $n$ on the chemical potential $\mu$  
obtained by the MPS methods used here to those obtained by 
numerically exact CT-QMC methods \cp{werner09}. 
The plateaus in $n(\mu)$ are the Mott insulating regimes of the phase diagram. 
The agreement is very good in general, confirming the reliability of our new procedure 
even with only three bath sites per correlated site. This leads to an extremely cheap computation, 
for which a single iteration of the DMFT loop took about \SI{30}{min} 
on two \SI{2.8}{GHz} cores (see \appref{secConv} for more details). 

\begin{figure}
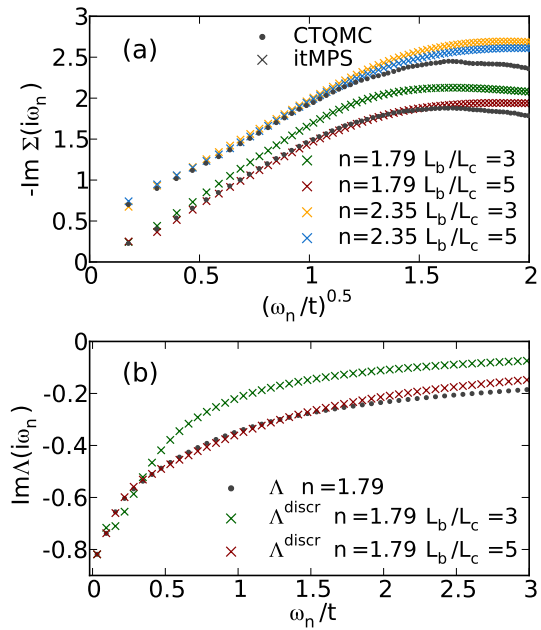

\ig{\numB}{pdf}{\figdir 00b_3band_SelfE}
\ig{\numB}{pdf}{\figdir 00d_3band_bathFit}
\caption{(Color online) 
Imaginary part of Matsubara axis self energy  $\Sigma$  (panel (a))  and imaginary part of hybridization function $\Lambda$ (panel (b)) for  densities shown obtained from converged itMPS solution of single-site DMFT for three-band Hubbard-Kanamori model \eqref{Hhubkan} for $U=8t$ and $J=U/6$.  Crosses (color online) represent itMPS data and black circles depict CTQMC data from Figure 3 of \oc{werner08}, computed at inverse temperature $\beta = 100/t$. We choose all parameters as described in the caption of \figref{fig3bdMIT}, in particular, for the bath fitting \eqref{costFunc}, we use $\beta\tl{eff}=100/t$, $\w_c=6t$ and $\alpha=1$. 
Choosing $\alpha=1$ enforces agreement for low frequencies at the price of 
disagreement at high frequencies, which is observed in both panels (a) and (b).
In panel (b), $\Lambda$ denotes the hybridization function that is fitted with the hybridization function $\Lambda\th{discr}$
of the discrete impurity model.
}
\label{fig3bdSpinFreeze}
\end{figure}

In panel (a) of  \figref{fig3bdSpinFreeze} we show a more stringent test, 
namely the dependence of the self energy on Matsubara frequency, 
in a parameter regime where the self energy was previously found \cite{werner08} 
to exhibit an anomalous $\omega^\frac{1}{2}$ frequency dependence  
and (in some regimes) a non-zero intercept as $\omega \rightarrow 0$. 
These phenomena are associated with a spin-freezing transition \cite{werner08,werner09}.

Panel (a) of Figure \ref{fig3bdSpinFreeze}  shows that the known low frequency $\omega \lesssim t$ self energy is already accurately reproduced even for the computationally inexpensive choice of  $L_b/L_c=3$, although one observes deviations for the high-frequency behavior. The deviations at high frequency decrease as the number of bath sites is increased, although full convergence at all frequencies has not been demonstrated.  Panel (b) of \figref{fig3bdSpinFreeze} shows that the deviations are linked to the impossibility of fitting the hybridization function equally well for all frequencies using only a small number of bath sites. The large deviations at high frequencies are due to the 
choice $\alpha=1$ in \eqref{costFunc}, which enforces good agreement for low frequencies and 
allowed to successfully reproduce the MIT transition in \figref{fig3bdMIT}. Increasing the number of bath 
sites to  $L_b/L_c=5$ leads to a  much better approximation of the hybridization function also 
for high frequencies, with concomitant improvement in the self-energy (\figref{fig3bdSpinFreeze}(a)).
\subsection{Three-band model in two-site DCA}

We now present results obtained using a two-site DCA approximation to   the three-band model of the previous section. For this problem there are no low-temperature results available in the literature. The size of the problem is beyond the scope of standard ED. The truncated configuration interaction (CI) impurity solver \cp{knizia12} allows to access a relatively high number of bath sites but is limited in the number of correlated sites: \eg in \oc{lin13ii}, a problem with $L_c=3$ and $L_b=30$ was computed, and in \oc{go15}, one with $L_c=4$ and $L_b=20$. The three-band two-site DCA though has $L_c=6$ correlated sites and it remains to be seen whether this is in reach for the CI solver. The problem is also challenging for standard CTQMC. Recent technical improvements on mitigating the sign problem \cp{nomura14} enabled \oc{nomura15} to treat this model at the  temperature of $T=0.025D$ with $D$ the half bandwidth, which required large computational resources. This temperature is relatively high as the authors stated that in the study of a simpler two-band two-site model, where they reached $T=0.0125D$, it was ``intractable'' to reach low enough temperatures to clarify whether bad metal/spin freezing behavior was intrinsic or not \cp{nomura14}.

We study  the model on the two-dimensional square lattice, i.e.~using $\varepsilon_k^{ab}=-2t\left(\cos k_x + \cos k_y\right)\delta^{ab}$.  We use the momentum patching of \oc{ferrero09}; this definition is also used in the single-band computations of \figsref{figImagTimeEvol} and \ref{figDCA}. We note that this model is not directly relevant to layered materials where the $t_{2g}$ orbitals are relevant, because in the physical situation the two dimensionality will break the three-fold orbital degeneracy. However the system is well defined as a theoretical model and useful to demonstrate the power of our methods.

As is the case for the CI method, the DMRG method used here is easily able to treat a large number of bath sites if the number of correlated sites is small: for $L_c=1$, DMRG has already often be proven to treat $L_b>120$ bath sites, and for $L_c=2$, $L_b>80$ is easily accessible \cp{wolf14,wolf14i}. However for more correlated sites, the number of bath sites that can be added at given computational cost decreases. For $L_c=6$, we use $L_b=18$, \ie $L_b/L_c=3$, which we have shown to be sufficient to produce reliable results in the single-site case without requiring  overly large computational resources (computation time of several hours per DMFT iteration on two cores).

\begin{figure}
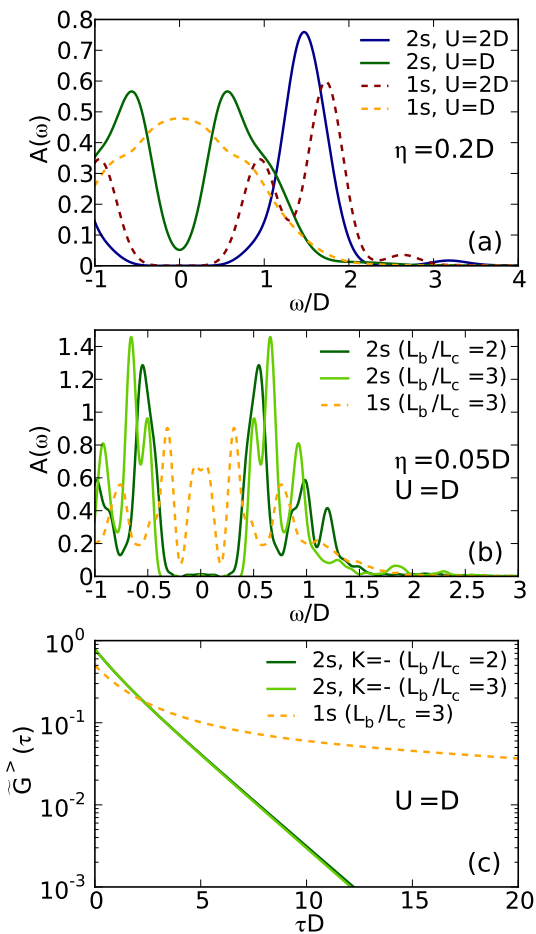

\ig{\numB}{pdf}{\figdir 03d_dca3bd_gap}
\ig{\numB}{pdf}{\figdir 03d_dca3bd_gapeta0.05}
\ig{\numB}{pdf}{\figdir 03d_dca3bd_gapimagtime}
\caption{(Color online) 
Comparison of results obtained using MPS methods with for $L_b/L_c=3$ for  single site (1s) and two-site (2s) DMFT approximations to the Hubbard-Kanamori model \eqref{HhubkanLat} on the two-dimensional square lattice with half  bandwidth $D=4t$, $\varepsilon_k^{ab}=-2t\left(\cos k_x + \cos k_y\right)\delta^{ab}$, $U'=U-2J$,  $J=U/4$ and $n=3$ ($\mu=5U/2-5J$),  that is, in the particle-hole symmetric case. Panel (a): spectral functions for  broadening $\eta=0.2D$; panel (b) broadening $\eta=0.05D$. In panel (c), we show the imaginary-time evolution of $\wt G^>(\tau)$  as defined in \eqref{Ggtrlssimag}, confirming by comparison to a calculation for a smaller bath $L_b/L_c=2$ that this quantity has been converged with respect to the bath size. The maximal bond dimension for the ground state search was $m=1000$.}
\label{fig3bdDCA}
\end{figure}

We have tested the two-site calculation by converging the DMFT loop for the three-band Hubbard model \Eqref{Hhubkan} with  $U'$ and $J=0$ and comparing the results with a corresponding two-site single-band DCA. Perfect agreement is obtained (not shown).  Non-zero values of $U'$ and $J$ create additional entanglement and make computations more costly. It is then a decisive question whether a real-space or a momentum-space representation of the impurity-cluster is less entangled. We discuss this in \appref{secBasis} finding that for the single-band Hubbard interaction, both representations yield similar entanglement, whereas for the Hubbard-Kanamori interaction, the real-space representation is much less entangled. Computational cost is therefore tremendously reduced by using the real-space representation, which comes with an off-diagonal hybridization function. This is the opposite behavior as observed for QMC, where the off-diagonal hybridization function creates a severe sign problem. We further note that in the real-space representation, strong interactions yield a less and less entangled impurity problem, as electrons become more and more localized.

We now present results for the more physically relevant case  $U'=U-2J$ with $J=U/4$. For these parameters, at half filling the critical interaction for the MIT in the single-site DMFT approximation is $U_c\simeq1.3D$ \cp{medici11}.
\figref{fig3bdDCA}(a) shows that our results are consistent with this estimate: the  dashed lines depict the single-site (1s) results, showing  a metallic solution (spectral function non-zero at $\omega=0$) for $U=D$, and an insulating solution (spectral function zero at $\omega=0$)  for $U=2D$.  In the two-site (2s) DCA (solid lines), by contrast, the critical value $U_c$  for the MIT is lowered. Even at  $U=D$ the $\omega=0$ spectral function is zero (the small non-zero value in panel (a) is an effect of broadening, as seen in panel (b)). The different nature of the metallic and insulating solutions is also  visible on the imaginary axis in the different nature of the decay of the imaginary-time Green's function. This is plotted in \figref{fig3bdDCA}(c) for $U=D$; clearly, a power-law decay is observed for the metallic solution obtained in the single-site DMFT, whereas an exponential decay is obtained for the insulating solution obtained within the two-site DCA.

\section{Conclusion}
\label{secCon}

This paper introduces an imaginary-time MPS (itMPS) solver for DMFT and shows that it can treat complex models, not easily accessible with other methods, at modest computational cost. This development establishes DMRG as a flexible low-coast impurity solver for realistic problems, such  as encountered in the study of strongly-correlated materials. The crucial advance stems from the fact that imaginary-time evolution does not create entanglement, and hence allows to compute Green's functions numerically exactly, provided a ground state calculation is feasible. 

The method  can be improved in many ways. In particular, different representations of the impurity problem exhibit different degrees of entanglement, so optimizing the representation of the impurity problem is a promising route. Ideas from ED approaches for constructing relevant subspaces \cp{lu14,zgid12,lin13i,lin13ii} of the  Hilbert space may lead to further improvements.  Such techniques have been successfully combined with MPS \cp{ma13}.  Another route to reduce computational effort and by that reach even more complex models could consist in performing computations for the reduced dynamics of the impurity \cp{cohen13ii}, or making use of extremely cheap algorithms for computing the Green's function at elevated temperatures \cp{savostyanov14}.  Finally, we note that using MPS as an impurity solver makes accessing entanglement as a quantity for understanding the properties of the embedded impurity cluster very easily accessible. Proposals in this direction have been made  for cellular DMFT \cp{udagawa15} and for impurity models  generally \cp{lee15}.

\section{Acknowledgements}

FAW thanks G.~K.-L.~Chan for stressing the relevance of 
converging the DMFT loop on the imaginary-frequency axis,
and N.-O. Linden for helpful discussions. AJM and US acknowledge the hospitality of the Aspen Center for Physics  NSF Grant 1066293 during the inception of this work. 
FAW and US acknowledge funding by \href{https://for1807.physik.uni-wuerzburg.de/}{FOR1807} of the DFG. AJM and AG were supported by the US Department of Energy under grant ER-046169.
\appendix

\section{Further technical details}
\label{secTech}
\subsection{Ground state optimization}
\label{secGS}

The main challenge in solving the ground state problem 
of a typical cluster-bath Hamiltonian as encountered
in DMFT, stems from the fact that DMRG is a variational 
procedure that is initialized with a random state,
which is then optimized locally. A local optimization
procedure is slow when optimizing a global 
energy landscape. In addition, the local optimization 
is prone to getting stuck in local minima, if no
``perturbation steps that mix symmetry sectors" are applied. 
The standard perturbation techniques for single-site DMRG \cp{white05,dolgov14,hubig15}
rely on ``perturbation terms" that are produced by contracting
the Hamiltonian with the MPS. If the Hamiltonian itself
does not contain terms that mix the symmetry sector,
these methods do not work. 

\begin{figure}
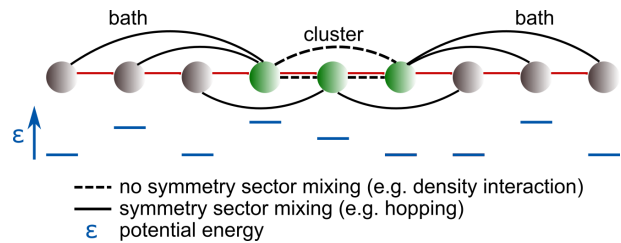

\ig{0.45}{png}{sketch_dmft_ham}
\caption{(Color online) 
Sketch of a typical cluster-bath Hamiltonian ($L_c=3$, $L_b=6$) 
when it's mapped to a one-dimensional chain.  
Dashed lines depict couplings that do not mix symmetry
sectors, and solid lines depict couplings that mix symmetry 
sectors. 
}
\label{figHam}
\end{figure}

A typical cluster-bath Hamiltonian has both features,
a global variation of the potential energy and parts 
that are not connected with symmetry mixing
terms, such as in the three-band Hubbard Kanamori model
at $J=0$. This situation is sketched in \figref{figHam}.

In \oc{wolf14}, the models under study allowed to solve this
problem using the non-interacting solution. For the general models
studied in the present paper, an unbiased numerical technique has
to be employed. What we do in practice, is to first find the ground
state of a system with additional symmetry-mixing couplings 
(denoted as red solid lines in \figref{figHam})
that are then adiabatically switched off. In practice, we sweep
5 to 10 times with additional hoppings of 10\%~magnitude of the physical hoppings,
and another 5 to 10 times with additional hoppings of 1\%~magnitude.
After these preliminary sweeps, the quantum number (\eg particle number)
distribution has globally converged, and we can continue with converging 
the ground state of the exact Hamiltonian.

\subsection{Convergence of DMFT iteration}
\label{secConv}

The calculations for the three-band single-site DMFT in \secref{sec3bd1s}   
are only trivially parallelized using one core to compute the imaginary  
time evolution of each the particle ($>$) and the 
hole ($<$) Green's functions $\wt G_{a}^{\gtrless}(\tau)$.

\begin{figure}
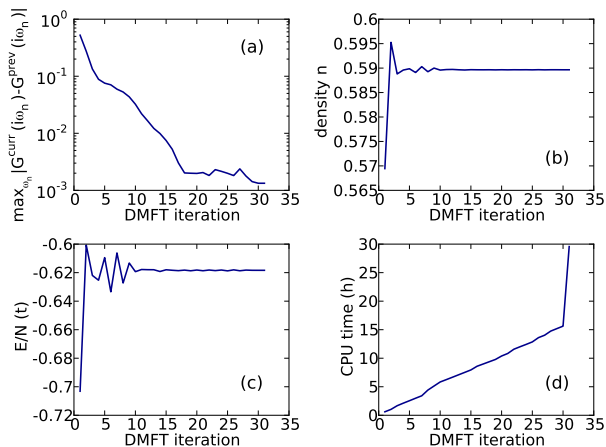

\ig{\numC}{pdf}{\figdir 00c_3band_convergence}
\ig{\numC}{pdf}{\figdir 00c_3band_convdens}
\ig{\numC}{pdf}{\figdir 00c_3band_convenergy}
\ig{\numC}{pdf}{\figdir 00c_3band_cput}
\caption{(Color online) 
Single-site DMFT for three-band Hubbard-Kanamori model
as studied in \figref{fig3bdSpinFreeze}. Here for 
the case $n=1.77$ ($\mu=5.0$) and $L_b/L_c=3$. To obtain
the solution for $n=1.79$ as shown in \figref{fig3bdSpinFreeze}, 
we chose $\mu=5.1$ and started from the $n=1.77$ solution.  
Panel (a): Convergence
of Matsubara Green's function in the DMFT loop, starting from the non-interacting
solution. 
Panel (b): Convergence of the density in the DMFT loop.
Panel (c): Convergence of the ground state energy 
per particle in the DMFT loop.
Panel (d): Computation time. An iteration on the Matsubara 
axis takes about \SI{30}{min}. The final real-axis
computation (iteration 31) is considerably more expensive,
but can still be optimized.
}
\label{figConvLoop}
\end{figure}

In \figref{figConvLoop}, we show the
convergence DMFT loop for the single-site DMFT for
the three-band Hubbard-Kanamori model
as studied in \figref{fig3bdSpinFreeze}. 
\figref{figConvLoop}(a) shows the convergence
of the Matsubara Green's function down to a precision of 
$10^{-3}$.  
\figsref{figConvLoop}(b) and (c) show the convergence of 
the density and of the ground state energy 
per particle, respectively.
\figref{figConvLoop}(d) shows the computation time. An iteration on the Matsubara 
axis takes about \SI{30}{min}. The final real-axis
computation (iteration 31) is considerably more expensive, 
but can still be optimized.

\section{Least-entangled representation and off-diagonal hybridization functions}
\label{secBasis}

\subsection{Geometry and general considerations}

In \oc{wolf14i}, some of us showed that the star geometry
of the impurity problem can have substantially 
lower entanglement than its chain geometry.
In the star geometry, DMRG profits from the small entanglement 
of the almost occupied states with low potential energy with
the almost unoccupied states with high potential energy.
A high weight for the superposition of a low- with a high-energy
state is physically irrelevant. In the star geometry, DMRG
is able to eliminate these superpositions as potential energy is separated locally,
i.e.~in the same basis in which DMRG optimizes the reduced
density matrix in order to discard irrelevant contributions.
In principle, as mentioned in \appref{secSpec}, 
ideas from basis selective approaches
in exact diagonalization are a different method to 
account for the fact that many states in the Hilbert space have a 
negligible weight for the computation of the Green's function 
and only few physically relevant states occupy 
a small fraction of the Hilbert space.
Among these are the truncated configuration
interaction (CI) \cp{knizia12,lin13i,lin13ii,go15}, the basis-selective ED \cp{lu14}
or the coupled cluster methods in quantum chemistry.
As these methods can be combined with DMRG \cp{ma13},
they might be a further route to construct efficient
representations of the impurity-cluster problem

In the present paper, the question of the least entangled 
representation of the impurity problem shall be restricted to the question
of which basis to choose in a DCA calculation. This is 
of high relevance also in another context: In the real-space
representation, the hybridization function becomes off-diagonal.
For CTQMC, this generates a sign problem. In our approach, this
doesn't affect computational cost much in the single-band 
Hubbard model. It even leads to a tremendous reduction 
of computational cost for the three-band Hubbard Kanamori interaction.

\subsection{DCA in momentum or real space}
\label{secDCArs}

The complexity of the interaction 
determines whether the real- or the momentum-space representation 
of the cluster-bath Hamiltonian  
is less entangled.
In real space, the interaction has a simple form, 
but the hybridization function has 
off-diagonal contributions, which result in additional couplings
of cluster and bath sites. In momentum space, the hybridization
function is diagonal but the interaction becomes off-diagonal. 
The additional couplings induced by that depend 
on the complexity of the interaction.

Let us be more concrete. For the two-site case, 
the discrete Fourier transform yields the even
and odd superposition of the real-space cluster.
\eq{
\wt d_{1}\dag & = \frac{1}{\sqrt{2}} (d_{1}\dag + d_{2}\dag) \\
\wt d_{2}\dag & = \frac{1}{\sqrt{2}} (d_{1}\dag - d_{2}\dag)
}
where the index of $\wt d_{K}\dag$ labels momentum
patches $K$ and the index of $d_{i}\dag$ labels real space
cluster sites $i$. There might be further indices labeling
spin or orbital.

In real-space, the hybridization function has the form
\eq{
\Lambda_{ij}(z) = \sum_{l=1}^{L_b} \frac{V_{il}^*V_{jl}}{z-\ve_l}, 
}
where the symmetry of the real-space cluster imposes 
$\Lambda_{ij}(z) = \Lambda_{ji}(z)$. 
In momentum space, the hybridization function is diagonal
\eq{
\wt\Lambda_{K}(z) = \sum_{l=l_K=1}^{L_b'} \frac{\wt V_{Kl}^*\wt V_{Kl}}{z-\wt\ve_{Kl}}
}
and symmetry is reflected in the reduced number 
of bath sites per patch  $L_b'=L_b/L_c$,
where $L_c=2$ is the number of momentum patches.

We choose to use the momentum representation for the bath
discretization, as was done for the real-axis in \oc{wolf14}.
While on the real-frequency axis this is the only viable option, 
the bath fitting on the imaginary-frequency axis 
via \eqref{costFunc} is possible also for the 
off-diagonal real-space case. In real space, \eg, particle hole 
symmetry can be easily imposed in the fitting procedure,
while this is not possible in momentum space.

Given the parameters of the momentum
space representation obtained by performing 
a bath fit via \eqref{costFunc},
we define the parameters of the equivalent 
real-space representation as follows:
In momentum space,
bath parameters are indexed by $l_K=1,...,L_b'$, $L_b'=L_b/L_c$
and in real space, bath parameters are indexed 
by $l=1,...,L_b$, then 
\eq{
\ve_l & = \wt\ve_{1,l_1=l} \tx{ for } l = 1,...,L_b' \non\\
\ve_l & = \wt\ve_{2,l_2=l-L_b'} \tx{ for } l = L_b'+1,...,L_b \non\\
V_{1l} & = V_{2l}  = \tfrac{1}{\sqrt{2}}\wt V_{1,l_1=l} \tx{ for } l = 1,...,L_b' \non\\
V_{1l} & = -V_{2l}  = \tfrac{1}{\sqrt{2}}\wt V_{2,l_2=l-L_b'} \tx{ for } l = L_b'+1,..., L_b.
}

Whereas the momentum-space Hamiltonian has  $L_b$  non-zero 
couplings $V_{Kl_K}$,
the real space Hamiltonian has $L_c\times L_b$ 
couplings $V_{il}$. On the other hand,
the interaction part generates $L_c\times(L_c-1)$
additional non-local couplings in the momentum-space
representation as compared to the real-space Hamiltonian.

From this one could naively expect that 
the real-space representation 
is less entangled if $L_c\times(L_c-1)>L_c\times L_b$.
Numerical experiments show
that the real-space representation is much more favorable
than this estimate. 
For a single-band Hubbard model, we find about the 
same entanglement in the real space and the momentum
space representation, with slight advantages for 
momentum space. In the three-band Hubbard Kanamori model,
the real-space representation is considerably less
entangled and leads to a tremendous reduction 
of computational cost. In particular, we were \tit{not} able 
to obtain the results of \figref{fig3bdDCA} in the momentum space 
representation when using $L_b/L_c=3$, only
for $L_b/L_c=2$ but then at much higher
computational cost.

\section{Green's functions from matrix product states}
\label{secSpec}

Even though the following discussion is not \tit{needed} to set up
the imaginary-time real-time impurity solver, it is of general
interest in this context and might stimulate further advancements.

\begin{figure}
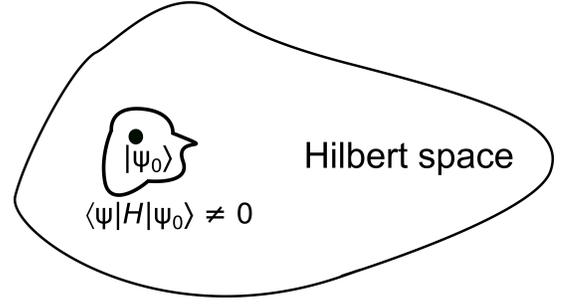

\ig{\numB}{png}{fig_subspace}
\caption{(Color online) 
Single particle excitation $\ket{\psi_0}$ of the ground state $\ket{E_0}$
and the subspace 
$\mathcal{N} = \{\ket{\psi}\,|\,\bra{\psi}H\ket{\psi_0} \neq 0\}$
of the Hilbert space
that is relevant for the computation
of a single-particle spectral function
of the form $\bra{\psi_0}\delta(\omega-H)\ket{\psi_0}$.
The single particle excitation is very lowly entangled, the subspace
is more strongly entangled, but still in general more lowly entangled
than the rest of the Hilbert space.
}
\label{figSubspace}
\end{figure}

A computation of $A(\w) = \bra{\psi_0}\delta(\w - (H-E_0)) \ket{\psi_0}$ 
via a computation of \tit{all} eigen states of $H$ 
is extremely redundant as only a tiny 
neighborhood $\mathcal{N} = \{\ket{\psi} \big| \bra{\psi}H\ket{\psi_0}\neq 0\}$ 
of a the single-particle excitation $\ket{\psi_0}$ contributes in
the sum (inserting identities $\sum_n \ket{E_n}\bra{E_n}$) in $A(\w)$. 
In ED, this is exploited by systematically 
constructing the subspace $\mathcal{N}$ by 
spanning it using particle-hole excitations \cp{lu14,knizia12},
which might also be a viable route for further 
developments within DMRG \cp{ma13}.
In DMRG, one needs to make a statement about the 
entanglement of the states in the subspace $\mathcal{N}$: 
one might note that these are in general more strongly entangled
than the single-particle excitation $\ket{\psi_0}$, but should still be much less
entangled than the rest of the Hilbertspace.
This is illustrated in the sketch \figref{figSubspace}.

In \oc{wolf15}, some of us argued
that expanding the spectral function in 
a family of orthogonal functions is a natural way to construct
a basis for $\mathcal{N}$, starting from the lowly entangled
$\ket{\psi_0}$ and successively increasing entanglement 
of states and thereby computational complexity in a sequence
of basis states $\ket{\psi_n}$. 
\oc{wolf15} discussed the expansion of $A(\w)$ in 
Chebyshev polynomials $T_n(\tfrac{\w}{a})=\arccos(n\cos(\tfrac{\w}{a}))$, which 
are orthogonal with respect to 
an inner product weighted by $w(x)=\sqrt{1-x^2}$ \cp{weisse06},  
and in the plane waves $\exp(i\w \tfrac{n}{a})$ (orthogonal with weight function $w(x)=1$), 
where the energy $a$ is chosen larger than the support of $A(\w)$. 
The associated generated sequences of basis states then are
\eq{
\ket{\psi_n\th{che}} 
& = 2(\tfrac{H-E_0}{a} + b)\ket{\psi_{n-1}\th{che}} - \ket{\psi_{n-2}\th{che}}, ~ b\in[-1,1], \non\\
\ket{\psi_n\th{time}} 
& = \exp(-i(H-E_0)\tfrac{n}{a})\ket{\psi_0},
}
and have different entanglement properties. 
The states $\ket{\psi_n\th{time}}$ associated with time evolution
are in general less entangled than the states $\ket{\psi_n\th{che}}$ associated
with the Chebyshev recursion \cp{wolf15}.  
This is due to the observation that error accumulation in the Chebyshev recursion 
is worse conditioned than in time propagation \cp{wolf15}, which necessitates
to keep the error in a single step of the Chebyshev recursion much smaller than in 
the equivalent time evolution step, which in turn requires to use higher bond dimensions 
in the Chebyshev recursion making it less efficient. 
In addition to the statements of \oc{wolf15}, we note here
that the sequence produced by the Lanczos algorithm,
\eq{
\ket{\psi_n\th{lan}} = H-\a_n \ket{\psi_{n-1}\th{lan}} - \ket{\psi_{n-2}\th{lan}}, ~ \a_n,\b_n \in \mathbb{R}
}
can be associated with an expansion of the spectral function
in polynomials that are orthogonal with respect to an inner
product weighted by $w(x) = A(x)$ \cp{gautschi05}. 
This is very efficient but numerically unstable.

In contrast to the previous methods, which generate 
an increasingly complex basis when determining 
the spectral function to a higher and higher precision, 
correction-vector DMRG aims to optimize a state 
in frequency space, which \tit{a priori}
contains contributions that have undergone an infinitely
long time evolution. As time evolution creates
entanglement, these states are much too strongly entangled
for an efficient treatment. 
They are ``far away'' from the controlled, lowly entangled
single-particle excitation $\ket{\psi_0}$. 
In order to still perform a meaningful computation in 
frequency space, one introduces a so-called (Lorentzian) broadening 
parameter $\eta$ that damps out contributions 
from an infinite time evolution. One does then not 
obtain the exact spectral function but a broadened version as in \Eqref{GretBroad}.
The broadening parameter has to be guessed 
\tit{a priori}: If it is chosen too small, 
high entanglement prevents convergence of the calculation. 
If it is chosen too large,
one will be far from the exact version of the spectral function. 
In the expansion methods discussed above, by contrast, 
one can stop the computation simply when it becomes too costly. 
If one has not recovered the exact $A(\w)$ at this point, 
a broadened version can be systematically constructed 
with an \tit{a posteriori} determined $\eta$ as in \Eqref{GretBroad}.

\input{imagmpsdmft.bbl}
%\bibliography{lit}

\end{document}

%% file: latexCommandDef.tex
\def\dag{^{\dagger}}

\newcommand{\bra}[1]{\langle #1 \vert} 
\newcommand{\ket}[1]{\vert #1 \rangle} 
\def\a{\alpha}
\def\b{\beta}

\def\ve{\varepsilon}

\def\w{\omega}

\newcommand{\appref}[1]{Appendix \ref{#1}}

\def\cp{\citep}

\def\ct{\citet}

\def\eg{e.g.\ }
\def\Eqref{Eq.~\eqref}

\newcommand{\expec}[1]{\langle #1 \rangle}

\newcommand{\figref}[1]{Fig.~\ref{#1}}
\newcommand{\figsref}[1]{Figs.~\ref{#1}}

\def\ie{i.e.\ }
\newcommand{\ig}[2]{\includegraphics[width=#1\tw,type=#2,ext=.#2,read=.#2]}

\def\lb{\label}

\def\non{\nonumber}

\def\oc{Ref.~\onlinecite}

\def\ra{\rightarrow}

\newcommand{\secref}[1]{Sec.~\ref{#1}}

\renewcommand{\th}[1]{^\tx{#1}}

\def\tit{\textit}

\newcommand{\tl}[1]{_\tx{#1}}

\def\tw{\textwidth}
\def\tx{\text}

\def\wt{\widetilde}

\newcommand{\eq}[1]{\begin{align} #1 \end{align}}
\def\bar{\begin{array}}
\def\ear{\end{array}}

\def\bce{\begin{center}}
\def\ece{\end{center}}

\def\ben{\begin{enumerate}}
\def\een{\end{enumerate}}

\def\bfi{\begin{figure}[!tbh]\setcapindent{0em}\centering}
\newcommand{\bfiO}[1]{\begin{figure}[#1]\setcapindent{0em}\centering}
\def\efi{\end{figure}}

\def\bit{\begin{itemize}}
\def\eit{\end{itemize}}

\def\bqu{\begin{quote}}
\def\equ{\end{quote}}

\newcommand{\btblO}[2]{\begin{table}[#2]\setcapindent{0em}\centering\begin{minipage}{#1\tw}}
\newcommand{\btbl}[1]{\begin{table}[!tbh]\setcapindent{0em}\centering\begin{minipage}{#1\tw}}
\def\etbl{\end{minipage}\end{table}}

\def\btbr{\centering\vspace{.4em}\begin{tabular}}
\def\etbr{\end{tabular}}

\def\bseq{\begin{subequations}}
\def\eseq{\end{subequations}}

\def\bve{\begin{array}{l}}
\def\eve{\end{array}{l}}

\newcommand{\igS}[3]{%
\hspace{-.025\tw}\begin{minipage}[t]{#1\tw}\vspace{0em}\includegraphics[width=\linewidth,type=#2,ext=.#2,read=.#2]{#3}\end{minipage}%
\hspace{.03\tw}\begin{minipage}[t]{1\tw-#1\tw-.05\tw}}

\newcommand{\lbS}[1]{\lb{#1}\end{minipage}}

\makeatletter
  \newcommand{\myhref}[2]{\hyper@linkurl{#2}{#1}}
\makeatother

\makeatletter
\newcommand*\if@single[3]{%
  \setbox0\hbox{${\mathaccent"0362{#1}}^H$}%
  \setbox2\hbox{${\mathaccent"0362{\kern0pt#1}}^H$}%
  \ifdim\ht0=\ht2 #3\else #2\fi
  }
\newcommand*\rel@kern[1]{\kern#1\dimexpr\macc@kerna}
\newcommand*\widebar[1]{\@ifnextchar^{{\wide@bar{#1}{0}}}{\wide@bar{#1}{1}}}
\newcommand*\wide@bar[2]{\if@single{#1}{\wide@bar@{#1}{#2}{1}}{\wide@bar@{#1}{#2}{2}}}
\newcommand*\wide@bar@[3]{%
  \begingroup
  \def\mathaccent##1##2{%
    \if#32 \let\macc@nucleus\first@char \fi
    \setbox\z@\hbox{$\macc@style{\macc@nucleus}_{}$}%
    \setbox\tw@\hbox{$\macc@style{\macc@nucleus}{}_{}$}%
    \dimen@\wd\tw@
    \advance\dimen@-\wd\z@
    \divide\dimen@ 3
    \@tempdima\wd\tw@
    \advance\@tempdima-\scriptspace
    \divide\@tempdima 10
    \advance\dimen@-\@tempdima
    \ifdim\dimen@>\z@ \dimen@0pt\fi
    \rel@kern{0.6}\kern-\dimen@
    \if#31
      \overline{\rel@kern{-0.6}\kern\dimen@\macc@nucleus\rel@kern{0.4}\kern\dimen@}%
      \advance\dimen@0.4\dimexpr\macc@kerna
      \let\final@kern#2%
      \ifdim\dimen@<\z@ \let\final@kern1\fi
      \if\final@kern1 \kern-\dimen@\fi
    \else
      \overline{\rel@kern{-0.6}\kern\dimen@#1}%
    \fi
  }%
  \macc@depth\@ne
  \let\math@bgroup\@empty \let\math@egroup\macc@set@skewchar
  \mathsurround\z@ \frozen@everymath{\mathgroup\macc@group\relax}%
  \macc@set@skewchar\relax
  \let\mathaccentV\macc@nested@a
  \if#31
    \macc@nested@a\relax111{#1}%
  \else
    \def\gobble@till@marker##1\endmarker{}%
    \futurelet\first@char\gobble@till@marker#1\endmarker
    \ifcat\noexpand\first@char A\else
      \def\first@char{}%
    \fi
    \macc@nested@a\relax111{\first@char}%
  \fi
  \endgroup
}
\makeatother